\documentclass[twocolumn]{svjour3}
\pdfoutput=1
\usepackage{newtxtext,newtxmath}
\usepackage{graphicx}
\usepackage[colorlinks=true]{hyperref}
\usepackage{siunitx}
\usepackage[misc]{ifsym}
\usepackage{xcolor}
\usepackage{natbib}
\usepackage{makecell}
\usepackage{amsmath,amssymb}

\makeatletter
\if@twocolumn
  \renewcommand\normalsize{%
   \@setfontsize\normalsize\@xpt{12.5pt}%
   \abovedisplayskip=3 mm plus6pt minus 4pt
   \belowdisplayskip=3 mm plus6pt minus 4pt
   \abovedisplayshortskip=0.0 mm plus6pt
   \belowdisplayshortskip=2 mm plus4pt minus 4pt
   \let\@listi\@listI}%

  \renewcommand\small{%
   \@setfontsize\small{8.5pt}\@xpt
   \abovedisplayskip 8.5\p@ \@plus3\p@ \@minus4\p@
   \abovedisplayshortskip \z@ \@plus2\p@
   \belowdisplayshortskip 4\p@ \@plus2\p@ \@minus2\p@
   \def\@listi{\leftmargin\leftmargini
               \parsep 0\p@ \@plus1\p@ \@minus\p@
               \topsep 4\p@ \@plus2\p@ \@minus4\p@
               \itemsep0\p@}%
   \belowdisplayskip \abovedisplayskip}

\else
  \if@smallext
   \renewcommand\normalsize{%
   \@setfontsize\normalsize\@xpt\@xiipt
   \abovedisplayskip=3 mm plus6pt minus 4pt
   \belowdisplayskip=3 mm plus6pt minus 4pt
   \abovedisplayshortskip=0.0 mm plus6pt
   \belowdisplayshortskip=2 mm plus4pt minus 4pt
   \let\@listi\@listI}%

  \renewcommand\small{%
   \@setfontsize\small\@viiipt{9.5pt}%
   \abovedisplayskip 8.5\p@ \@plus3\p@ \@minus4\p@
   \abovedisplayshortskip \z@ \@plus2\p@
   \belowdisplayshortskip 4\p@ \@plus2\p@ \@minus2\p@
   \def\@listi{\leftmargin\leftmargini
               \parsep 0\p@ \@plus1\p@ \@minus\p@
               \topsep 4\p@ \@plus2\p@ \@minus4\p@
               \itemsep0\p@}%
   \belowdisplayskip \abovedisplayskip}
 \else
  \renewcommand\normalsize{%
   \@setfontsize\normalsize{9.5pt}{11.5pt}%
   \abovedisplayskip=3 mm plus6pt minus 4pt
   \belowdisplayskip=3 mm plus6pt minus 4pt
   \abovedisplayshortskip=0.0 mm plus6pt
   \belowdisplayshortskip=2 mm plus4pt minus 4pt
   \let\@listi\@listI}%

  \renewcommand\small{%
   \@setfontsize\small\@viiipt{9.25pt}%
   \abovedisplayskip 8.5\p@ \@plus3\p@ \@minus4\p@
   \abovedisplayshortskip \z@ \@plus2\p@
   \belowdisplayshortskip 4\p@ \@plus2\p@ \@minus2\p@
   \def\@listi{\leftmargin\leftmargini
               \parsep 0\p@ \@plus1\p@ \@minus\p@
               \topsep 4\p@ \@plus2\p@ \@minus4\p@
               \itemsep0\p@}%
   \belowdisplayskip \abovedisplayskip}
  \fi
\fi

\makeatother

\usepackage{microtype}
\journalname{Rheologica Acta}
\begin{document}

\title{Turning a yield-stress calcite suspension into a shear-thickening one by tuning inter-particle friction}
\author{James A. Richards \thanks{\Letter~~~James A. Richards \newline \hspace*{5mm}jamesrichards92@gmail.com} \and Rory E. O'Neill \and Wilson C. K. Poon}
\institute{SUPA, School of Physics and Astronomy, University of Edinburgh, James Clerk Maxwell Building, Peter Guthrie Tait Road, Edinburgh, EH9 3FD, United Kingdom}
\date{\today}
\maketitle

\begin{abstract}
We show that a suspension of non-Brownian calcite particles in glycerol-water mixtures can be tuned continuously from being a yield-stress suspension to a shear-thickening suspension--without a measurable yield stress--by the addition of various surfactants. We interpret our results within a recent theoretical framework that models the rheological effects of stress-dependent constraints on inter-particle motion. Bare calcite particle suspensions are found to have finite yield stresses. In these suspensions, frictional contacts that constrain inter-particle sliding form at an infinitesimal applied stress and remain thereafter, while adhesive bonds that constrain inter-particle rotation are broken as the applied stress increases. Adding surfactants reduces the yield stress of such suspensions. We show that, contrary to the case of surfactant added to colloidal suspensions, this effect in non-Brownian suspensions is attributable to the emergence of a finite onset stress for the formation of frictional contacts. Our data suggest that the magnitude of this onset stress is set by the strength of surfactant adsorption to the particle surfaces, which therefore constitutes a new design principle for using surfactants to tune the rheology of formulations consisting of suspensions of adhesive non-Brownian particulates. 

\keywords{Suspension \and Yield stress \and Shear thickening \and Rheology \and Calcite \and Dispersants}
\end{abstract}

\section*{\label{sec:intro}Introduction}

A transformation has recently occurred in our understanding of the rheology of  suspensions of hard non-Brownian (nB) particles (size $\gtrsim \SI{2}{\micro\meter}$) with repulsive interactions. Such suspensions shear thicken: their viscosity increases with applied shear rate~\citep{barnes1989shear}. It is now accepted that the stress-driven contact formation plays the dominant role~\citep{lin2015hydrodynamic, clavaud2017revealing, comtet2017pairwise}. Particles in mechanical contact cannot freely slide past one another due to Coulomb friction~\citep{seto2013discontinuous} or other mechanisms~\citep{hsu2018roughness,james2018interparticle}. This happens when the stabilising repulsive force between particles fails to keep them separated when the applied stress increases beyond a critical threshold, $\sigma^*$, the ``onset stress''. The close approach of particle surfaces switches on anti-sliding mechanisms~\citep{wilson2000viscosity,wyart2014discontinuous}. The additional particle motion needed on the local level to accommodate any given macroscopic strain leads to extra dissipation, so that the viscosity rises~\citep{lerner2012unified}. 

Most industrial nB suspensions are not purely repulsive. Typically, at high enough volume fraction, $\phi$, there exists a yield stress, $\sigma_y$, below which there is no flow. Above $\sigma_y$ the suspension shear thins and the viscosity decreases to a limiting plateau value. Practical examples span diverse sectors, from suspensions of mineral powders~\citep{zhou1995yield} and polymeric latices~\citep{heymann2002solid} to coal slurries~\citep{wildemuth1985new} and molten chocolate~\citep{blanco2019conching}, for which glass spheres with hydrophobic coating in water~\citep{brown2010generality} may function as a generic model system. In such applications, it is important to be able to `tune' the yield stress, for example in unset concrete. To allow pumping and filling of formwork, $\sigma_y$ must not be too high \citep{roussel2007rheology}; in contrast during 3D printing of concrete, too low a $\sigma_y$ and the unset concrete will not hold the desired shape before setting \citep{mechtcherine2020extrusion}. It is therefore important to understand the origin of the yield stress in nB suspensions. 

A finite suspension yield stress is typically traced back to residual van der Waals attraction between particles that are insufficiently stabilised~\citep{bonn2017yield}. Such particles can `bond' so that above some critical $\phi \ll \phi_{\rm rcp}$ they form a stress-bearing network. To flow, a finite stress $\sigma > \sigma_y$ must be applied to break the bonds and fluidise the suspension. A classic way to tune $\phi_y$ is by adding surfactants, variously known as dispersants, plasticisers or other sector-specific terms. These adsorb onto particle surfaces, increase the minimum separation, and so reducing attraction and $\sigma_y$. 

This explanation undoubtedly applies to colloids, where Brownian motion drives aggregation, and has been used to explain yield stress in nB suspensions~\citep{brown2010generality}. To see that something may be amiss in the latter case, consider an aqueous suspension of calcite particles, as found in toothpastes, paints and paper coatings. Later, we show that a $\phi \approx 0.5$ suspension of this kind with particle dimension $d \approx \SI{4}{\micro \metre}$ has $\sigma_y \approx \SI{D2}{\pascal}$ under steady shear. Dimensional analysis of the colloidal picture suggests that $\sigma_y \sim U/d^3$, with the energy scale $U$  set by the van der Waals interaction. Taking  $U \sim A d / 12 h$ for two spheres at surface separation $h$ with the Hamaker constant, $A$, of calcite in water~\citep{bergstrom1997hamaker}, we find $h \sim \SI{0.01}{\angstrom}$, far below the atomic scale. It is evident we are missing some vital physics. 

The missing physics is particle contact, which, if attractive forces are present, can prevent rolling below a critical torque when the contacts are pinned~\citep{heim1999adhesion,estrada2011identification}. Such adhesive contacts constrain inter-particle rolling, just as frictional contacts prevent sliding. The effect of sliding and rolling constraints acting independent or together on viscosity has been explored in a `constraint rheology' framework~\citep{guy2018constraint}, in which there are two critical stress scales, the onset stress for making frictional contact, $\sigma^*$, and the strength of an adhesive contact, $\sigma_a$. 

Below we review constraint rheology~\citep{guy2018constraint}, and then consider our data for a model calcite suspension in the context of constraint rheology. The finite yield stress in the bare-particle system can be `tuned away' by adding surfactants. Making sense of our observations using the constraint rheology framework allows us to interpret the role of different surfactants in such adhesive nB suspensions, which turns out to differ radically from how they act in Brownian suspensions. 

\section*{\label{model}Constraint rheology of suspensions}

\begin{figure}
\centering
\includegraphics{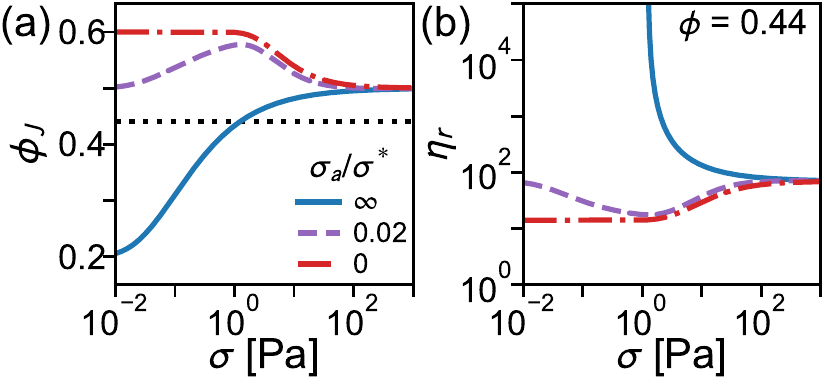}
\caption{Changing the ratio of the frictional onset stress to adhesive strength, $\theta = \sigma_a/\sigma^*$. (a)~Stress-dependent jamming point, $\phi_J(\sigma)$, with decreasing $\theta$, Eqs.~\ref{eq:f}-\ref{eq:phiJ2}. Lines: solid (blue), yield-stress ($\sigma_a\! =\! \SI{0.1}{\pascal}$, $\sigma^* \!=\! \SI{0}{\pascal}$); dashed (purple), thinning then thickening ($\sigma_a \!=\! \SI{0.1}{\pascal}$, $\sigma^* \!=\! \SI{5}{\pascal}$); dot-dashed (red), shear-thickening ($\sigma_a\! =\! \SI{0}{\pascal}$, $\sigma^*\! =\! \SI{5}{\pascal}$); and, dotted (black), example volume fraction, $\phi=0.44$. Other parameters: $\phi_{\rm rcp}\!=\!0.6$, $\phi_{\mu}\! = \!0.50$, $\phi_{\rm alp}\!=\!0.2$, $\beta\! = \!1$, $\kappa \! = \! 0.6$. (b)~Resultant flow curves at $\phi=0.44$ using Eq.~\ref{eq:KD} ($\ell=2$), $\sigma_a/\sigma^*$ as in (a)}
\label{fig:constraint}
\end{figure}

For suspensions of hard particles, the relative viscosity, $\eta_r$, is controlled by the proximity to the jamming volume fraction. The viscosity divergence at some jamming point, $\phi_J$, is captured by the form of~\citet{krieger1959mechanism},
\begin{equation}
    \label{eq:KD}
    \eta_{r} = \eta / \eta_{s} = \left[ 1 - \phi/\phi_J \right]^{-\ell},
\end{equation}
with $\ell \gtrsim$ 2~\citep{guy2015towards}. At $\phi \geq \phi_J$, the system shear jams and fractures under deformation~\citep{brown2014shear,dhar2020signature}. In the~\citet{wyart2014discontinuous} (WC) model for shear thickening, $\phi_J$ depends on the stress-dependent fraction of frictional contacts,
\begin{equation} 
    f(\sigma) = \exp[ -\left( \sigma^*/\sigma \right)^{\beta}],
\label{eq:f}
\end{equation}
which increases at the onset stress, $\sigma^*$, from $f(\sigma \! \ll \! \sigma^*)=0$ to $f(\sigma \! \gg \! \sigma^*)=1$, with a rapidity set by $\beta$. The increasing fraction of frictional contacts lowers the jamming point via
\begin{equation}
\phi_{J} = \phi_{\mu} f + \phi_{\rm rcp} (1-f),
\label{eq:phiJ1}
\end{equation}
from random close packing, $\phi_J(f\!=\!0) = \phi_{\rm rcp}$ at $\sigma \ll \sigma^*$, to a lower frictional jamming point, $\phi_J(f\!=\!1) = \phi_\mu$, at $\sigma \gg \sigma*$, whose value depends on inter-particle friction~\citep{silbert2010jamming}. For a suspension at a fixed $\phi < \phi_\mu$, a $\phi_J(\sigma)$ decreasing with stress gives a viscosity that increases from a low-shear plateau, $\eta_r^0$, to a high-shear plateau, $\eta_r^\infty$, Eq.~\ref{eq:KD}, as observed. 

Importantly, adhesive constraints that limit inter-particle rolling can also lower $\phi_J$~\citep{guy2018constraint,richards2020role}. These are broken above a critical torque, $M_a$, set by the attractive force between particles and a surface length scale that pins the contact~\citep{heim1999adhesion}. The fraction of adhesive contacts, $a$, decreases rapidly above a characteristic stress, $\sigma_a \sim M_a/d^3$, which we model by
\begin{equation}
\label{eq:a}
a(\sigma) = 1- \exp\left[ -\left( \sigma_a/\sigma \right)^{\kappa}\right],
\end{equation}
where $\kappa$ controls how rapidly $a$ decreases from $a(\sigma \! \ll \! \sigma_a)=1$ to $a(\sigma \! \gg \! \sigma_a)=0$. 

With adhesive constraints alone, jamming occurs at `adhesive close packing', $\phi_J(a\!=\!1,f\!=\!0)=\phi_{\rm acp}$. Since the number of constraints (two rolling degrees of freedom) is the same as that in a purely frictional system (two sliding degrees of freedom), we take $\phi_{\rm acp} = \phi_\mu$. If both constraints operate, jamming occurs at a lower concentration, `adhesive loose packing', $\phi_J(a\!=\!1,f\!=\!1)=\phi_{\rm alp} < \phi_{\rm acp}$. This critical volume fraction, which is possibly related to rigidity percolation~\citep{richards2020role}, is not yet precisely known; one simulation returns $\phi_{\rm alp} \approx 0.14$ \citep{liu2017equation}. The various jamming points are summarised in Table~\ref{tab:phiJ}. 

In any actual suspension, $0 \leq f\!\!\left(\sigma/\sigma^*\right)\!, \,a\!\left(\sigma/\sigma^*\right) \leq 1$, and $\phi_J(\sigma)$ depends on the degree to which frictional/adhesive contacts are formed/broken by the applied stress. We use a phenomenological {\it ansatz} to extend Eq.~\ref{eq:phiJ1}:
\begin{equation}
\begin{split}
\phi_{J} = &f \left[ \phi_{\rm alp} a + \phi_{\mu} (1-a)\right] +\\
&(1-f)\left[ \phi_{\rm acp} a + \phi_{\rm rcp} (1-a) \right].
\label{eq:phiJ2}
\end{split}
\end{equation}

\begin{table}
\centering
\begin{tabular}{|c|c|c|}
\hline
  \thead{Frac.~frictional\\contacts, $f$} & \thead{Frac.~adhesive\\contacts, $a$} & $\phi_J$\\
  \hline\hline
   0  & 0 & $\phi_{\rm rcp}$\\
   \hline
   1  & 0 & $\phi_\mu$ \\
   \hline
   0 & 1 & $\phi_{\rm acp} = \phi_\mu$\\
   \hline
   1 & 1 & $\phi_{\rm alp}$\\
   \hline
\end{tabular}
\caption{Table of jamming volume fractions in decreasing order. For quasi-monodisperse hard spheres, $\phi_{\rm rcp} \approx 0.64$ and $\phi_\mu \approx 0.55$ are well documented; one simulation suggests $\phi_{\rm alp} \approx 0.14$} \label{tab:phiJ}
\end{table}

The rheology clearly depends on the ratio $\theta = \sigma_a/\sigma^*$. In a `frictional suspension', $\theta \gg 1$ ($\sigma^* \ll \sigma_a $), $f=1$ at all accessible stresses and the flow is always frictional. Adhesion stabilises frictional contact networks, so that the system can jam and a yield stress develops at some rather low $\phi_{\rm alp} $. Increasing the stress on a jammed system releases proportionately more adhesive constraints, so that $\phi_J$ monotonically increases from $\phi_{\rm alp}$ to $\phi_{\mu}$. This monotonic $\phi_J(\sigma)$ produces a corresponding monotonic shear-thinning $\eta_r(\sigma)$, Eq.~\ref{eq:KD} and Fig.~\ref{fig:constraint}~[solid (blue)]. 

In a `lubricated suspension', $\theta \ll 1$ ($\sigma^* \gg \sigma_a$), increasing $\sigma$ beyond $\sigma_a$ rapidly releases rolling constraints while $f \approx 0$. Thus, $\phi_J$ increases from $\phi_{\rm acp}$ (which we take to be $= \phi_\mu$) towards $\phi_{\rm rcp}$, until $\sigma \rightarrow \sigma^*$ and frictional contacts start to form, whereupon $\phi_J$ decreases towards $\phi_\mu$. Such a non-monotonic $\phi_J(\sigma)$ gives rise to a corresponding non-monotonic $\eta_r(\sigma)$, Eq.~\ref{eq:KD} and Fig.~\ref{fig:constraint}~[dashed (purple)]. In this case, with no frictional contact network for adhesion to stabilise, the system cannot jam (and $\sigma_y = 0$) below $\phi_{\rm acp}$ ($=\phi_\mu$ for us)~\citep{richards2020role}. 

Within this framework, then, we may `tune' a suspension in the range $\phi_{\rm alp} < \phi < \phi_{\rm acp} = \phi_\mu$ from having a finite $\sigma_y$ to having essentially $\sigma_y \to 0$--a many orders of magnitude change--by engineering a transition from the $\theta \gg 1$ (frictional) regime to $\theta \ll 1$ (lubricated) regime. Below we show how to do this using surfactants in a calcite suspension. 

\section*{Materials and methods\label{sec:methods}}

\begin{figure}
\includegraphics[width=\columnwidth]{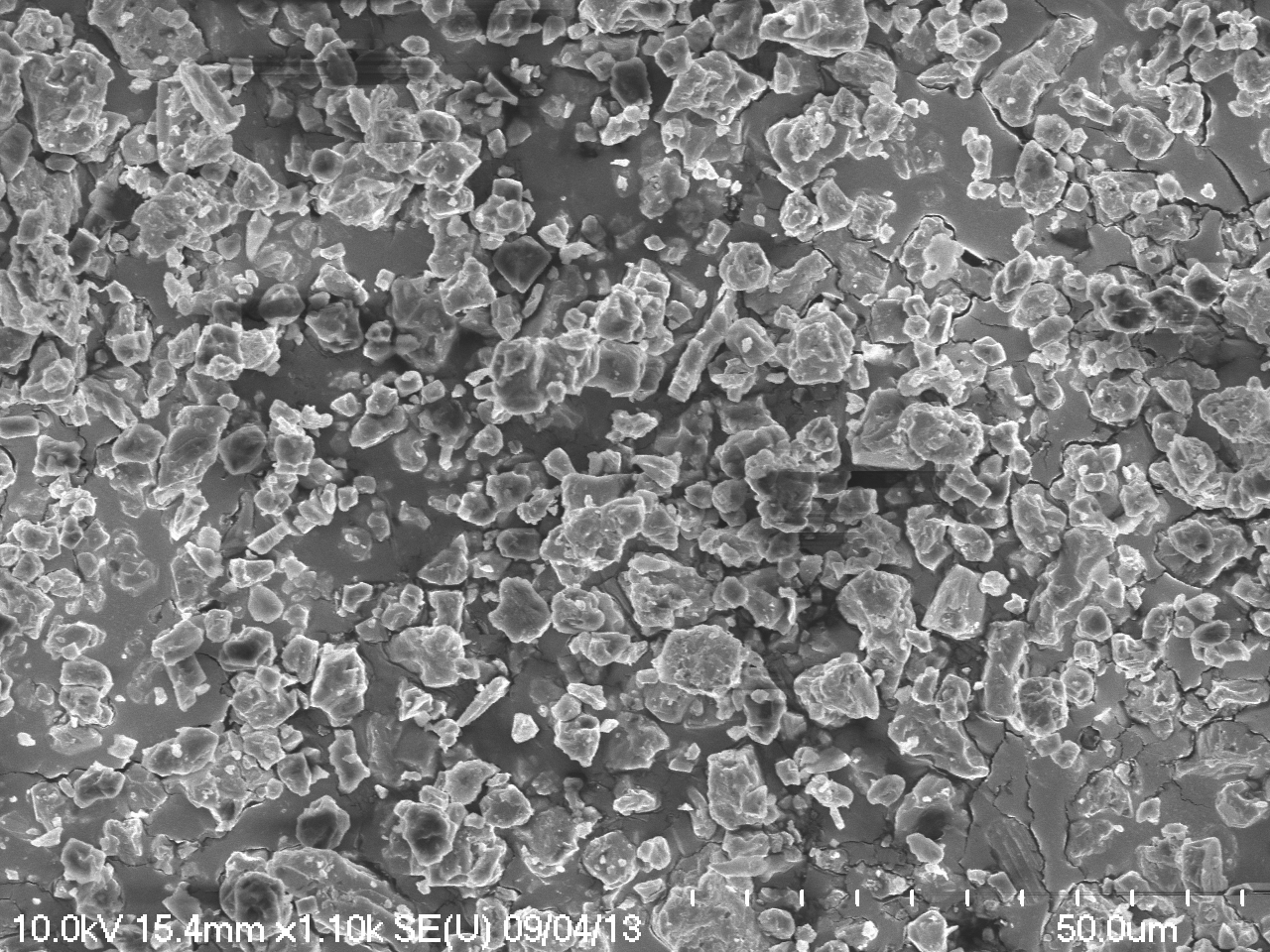}
\caption{Scanning electron micrograph of ground calcite, scale bar \SI{50}{\micro \metre}}
\label{fig:SEM}
\end{figure}

We studied ground calcium carbonate with a rhombohedral form [Eskal 500, \citet{eskal500data}, 99\% purity, density $\rho_p = \SI{2.7}{\gram \per \centi \metre^3}$, $d_{50} = \SI{4}{\micro\metre}$], Fig.~\ref{fig:SEM}. Ground calcite is widely used as a filler in aqueous coatings to improve abrasion resistance and finish, or as a cheap extender. Powder was dispersed in glycerol-water mixtures using vortex and then high-shear mixing at $\phi \leq 0.45$, or manual stirring at $\phi > 0.45$, until a smooth appearance was achieved. Using a glycerol-water mixture slows evaporation and sedimentation, both of which can prevent accurate rheology. The glycerol content was adjusted to access the maximum range of stresses in each sample ($\SI{0.1}{\pascal} \lesssim \sigma \lesssim \SI{400}{\pascal}$). We checked that varying the amount of glycerol did not strongly change the rheology or qualitatively influence our conclusions. 

We used three surfactants: polyacrylic acid (PAA); an alkyl-napthalene sulphonate condensate (ANS), Morwet D-425; and a polycarboxylate ether (PCE), Agrilan 755. PAA adsorbs to calcite forming a monolayer~\citep{eriksson2007calcite}. Our PAA was a linear \SI{5100}{\dalton} sodium salt (Sigma Aldrich) with a \SI{3}{\nano \metre} gyration radius~\citep{reith2002does}. Commercial ANS is a highly-polydisperse mixture of oligomers with branched and cross-linked chains~\citep{piotte1995characterization} often used as ``superplasticisers'' in self-compacting concrete~\citep{mehta1999advancements}. The PCE is a comb co-polymer of polyethylene glycol grafted to a methacrylate-methylmethacrylate backbone. Similar surfactants are known to stabilise calcite suspensions~\citep{bossis2017discontinuous}. 

Surfactants were dissolved into the glycerol-water mixture before powder incorporation at concentrations reported as weight percentages relative to the solid content (w/w\%). The main concentrations used were 0.05 w/w\% (PAA), 0.5 w/w\% (ANS) and 1.0 w/w\% (PCE), which were chosen as the concentration needed in each case before no further change was found in the rheology of a $\phi=0.44$ sample if more is added (see Appendix). 


Steady-state flow curves were measured using parallel plates (radius $R=\SI{20}{\milli \metre}$ and gap height $H=\SI{1}{\milli\metre}$) that were sandblasted or serrated to reduce slip. For suspensions without surfactant, controlled shear rate measurements were taken (TA Instruments ARES-G2, roughened plates). We report the rim shear rate, $\dot{\gamma} = \Omega R/ H$ from the applied angular velocity, $\Omega$; the stress, $\sigma = (\mathcal{T}/2\pi R ^3)(3+ {\rm d} \ln \mathcal{T}/ {\rm d}\ln \Omega)$ from the measured torque, $\mathcal{T}$; and hence the relative viscosity, $\eta_r = \sigma/(\dot{\gamma}\eta_s)$. 

Samples were pre-sheared at high stress to remove the loading history, before applying a single upsweep at 5 points per decade from the minimum shear rate, $\dot{\gamma}_{\min} = \SI{0.1}{\per \second}$, to sample fracture at $\sigma_{\max} \approx \SI{400}{\pascal}$. At each point the longer of $\gamma = 10$ or $t = \SI{10}{\second}$ was accumulated with an average of the steady state taken. The minimum shear rate is set by the longest experiment limited by sedimentation below $\sigma_{\min} = \Delta \rho g d \approx \SI{0.1}{\pascal}$. When strong shear thinning is observed we identify an experimental yield stress, $\sigma_y = \sigma(\dot{\gamma}_{\rm min})$. 

Measurements with surfactants used imposed stress (serrated plates, TA Instruments DHR-2 for ANS and PCE, AR-2000 for PAA). After a \SI{1}{\pascal} pre-shear, 10 points per decade were measured between $\sigma_{\min}$ and fracture or inertial sample ejection, always ensuring reversibility below fracture. At each point, measurement followed a \SI{5}{\second} equilibration, but the total time per point was adjusted between samples to maximise the averaging time while still avoiding sedimentation. For PAA and PCE the step time was \SI{10}{\second} for $\phi \leq 0.4$, \SI{20}{\second} for $0.45 \leq \phi \leq 0.49$ and \SI{30}{\second} for $\phi \geq 0.51$; for ANS a single step time of \SI{15}{\second} was used. For systems where the shear rate may decrease with stress (discontinuous shear thickening), we report the apparent stress, $\sigma_{\rm app} = 2\mathcal{T}/ \pi R^3$. 

\section*{\label{sec:results}Results}


\begin{figure}[t]
\includegraphics[width=\columnwidth]{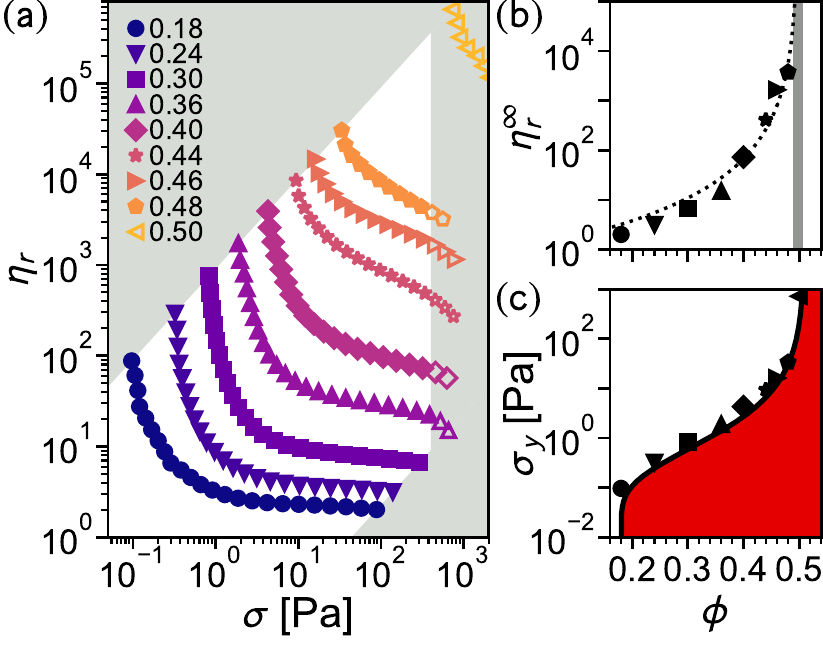}
\caption{Steady-state rheology of bare calcite suspensions. (a)~Relative viscosity vs stress, $\eta_r(\sigma)$, under imposed shear rate. Symbols, data at volume fraction, $\phi$ (legend). Shading (grey) outside measurable limits. (b)~High-shear relative viscosity, $\eta_r^{\infty}(\phi)$. Symbols, $\eta_r$ before fracture; dotted line, $\eta_r^{\infty} = (1- \phi/\phi_\mu)^{-\ell}$, $\phi_\mu = 0.50(1)$ [shading (grey)] and $l = 2.6(3)$ (c)~Yield stress, $\sigma_y$, vs volume fraction. Symbols, yield stress from minimum shear rate, $\sigma_y = \sigma(\dot{\gamma}_{\min} = \SI{0.1}{\per \second})$. Solid line, $\sigma_y$ with $\theta \to \infty$, from Eqs.~\ref{eq:a} and \ref{eq:phiJ2} ($f=1$ and $\phi_{\rm alp} =0.18$), taking $\sigma_a = \SI{0.6}{\pascal}$ and $\kappa = 0.6$. Shaded (red), jammed; and unshaded, flowing }
\label{fig:bareFC}
\end{figure}

Suspensions of bare calcite particles in an 85~wt.\% glycerol-water mixture ($\eta_s = \SI{110}{\milli \pascal \second}$) show strong shear thinning at all measured $\phi$, Fig.~\ref{fig:bareFC}(a), consistent with the presence of a yield stress below which flow ceases. At $\phi < 0.40$, the viscosity decreases towards a high-shear plateau, $\eta_r^{\infty}$. At higher $\phi$ any plateau value is obscured by sample fracture (open symbols). At $\phi=0.50$ all flow may be due to fracture. In all cases, we take $\eta_r^{\infty}$ to be the viscosity at the highest stress before fracture. 

The absence of shear thickening suggests that the system is always frictional. This is confirmed by fitting Eq.~\ref{eq:KD} to $\eta_r^{\infty}(\phi)$, giving divergence at $\phi_{\mu} = 0.50(1)$ [with $\ell = 2.6(3)$], Fig.~\ref{fig:bareFC}(b). With $\sigma_y \approx \sigma_{\min}$ at $\phi = 0.18$, it was not possible to measure a yield stress for lower $\phi$ without sedimentation; so, we take (as an upper bound) $\phi_{\rm alp} = 0.18$. Using an adhesive strength of $\sigma_a \approx \SI{0.6}{\pascal}$ (and $\kappa = 0.6$) in the constraint model of Eq.~\ref{eq:phiJ2} with $f = 1$ and Eq.~\ref{eq:a}  then predicts a $\phi$-dependence of $\sigma_y$ that accounts well for our observations, Fig.~\ref{fig:bareFC}(c). 

Physically, Fig.~\ref{fig:bareFC}(c) tells us that at low $\phi$ the suspension flows at all applied stresses. When $\phi$ reaches $0.18$, it becomes possible for adhesion to stabilise the frictional contact network enough for the system to jam. However, this adhesion-stabilised frictional network breaks above a finite applied stress, $\sigma_y$, and the system flows. As $\phi$ increases, the frictional network acquires additional stability, and more adhesive bonds need to be broken to fluidise the suspension: $\sigma_y$ increases. Eventually, upon reaching $\phi_\mu$, no adhesive stabilisation is needed - the frictional network is stable in its own right: the system is jammed at all stresses and $\sigma_y \to \infty$. 

\begin{figure}
\includegraphics[width=\columnwidth]{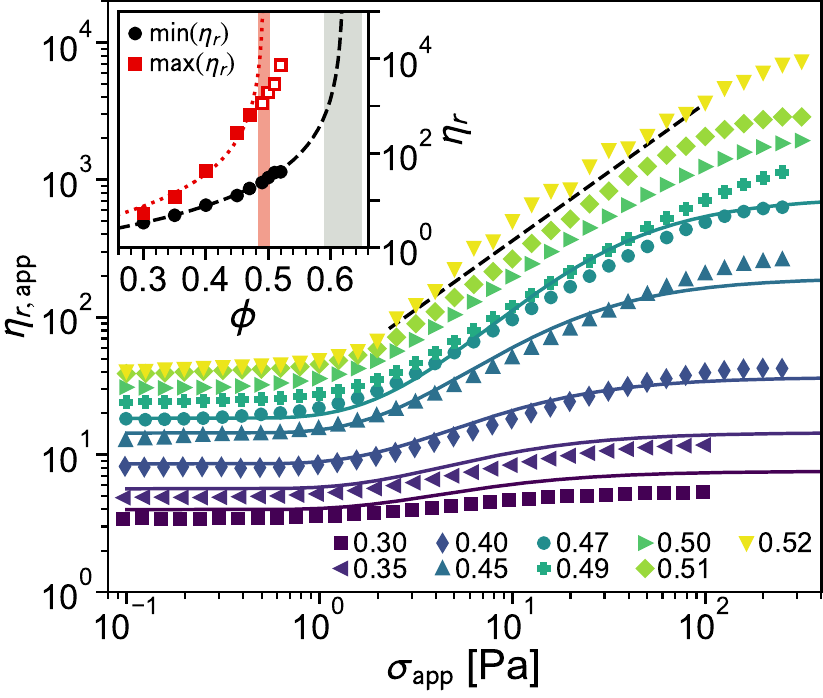}
\caption{Flow curves of polyacrylic acid (PAA) stabilised calcite suspensions. Symbols: apparent relative viscosity vs stress, $\eta_{r,{\rm app}}(\sigma_{\rm app})$, at volume fraction, $\phi$ (legend). Dashed (black) line, DST (slope = 1). Solid lines, WC model ($\theta \to 0$) fit to $\phi \leq 0.47$, onset stress $\sigma^* = \SI{3.0(2)}{\pascal}$ ($\beta=1.02$). Inset: plateau viscosity with volume fraction. Squares (red): high-shear viscosity from $\max(\eta_r)$; circles (black), low-shear viscosity from $\min(\eta_r)$. Dotted (red) line, fit of Eq.~\ref{eq:KD}, to find $\phi_{\rm J} = 0.49(1)$ [shading (red)] and $\ell=2.2(2)$. For $\min(\eta_r)$ a pre-factor, $A= 0.8(1)$, is included in Eq.~\ref{eq:KD} [dashed (black) line], to give $\phi_{\rm J} = 0.62(3)$ [shading (grey)]}
\label{fig:PAAfc}
\end{figure}

Thus, bare calcite particles in a glycerol-water mixture form a frictional suspension: little or no stress is needed to push the particles into frictional contact because $\sigma^* \rightarrow 0$. This onset stress can be made finite by introducing a repulsive barrier between particles, which then must be overcome to press particles into mechanical contact. This can be done, e.g., via electrostatic surface effects by modifying the dissolved ions~\citep{almahrouqi2017zeta}. We will tune the repulsive barrier sterically using surfactants. 


When PAA is added (70~wt.\% glycerol-water mixture, $\eta_s = \SI{22.5}{\milli \pascal \second}$), the suspension no longer shows any evidence of a finite yield stress, Fig.~\ref{fig:PAAfc}~(symbols). Instead, the viscosity rises from a low-shear to a high-shear plateau at increasing stress, with the magnitude of the effect increasing with volume fraction. This is classical shear thickening. 

In detail, at $\phi \leq 0.47$, suspensions continuously thicken to a high-shear plateau, but for $\phi > 0.47$ a plateau is not reached before fracture. At the highest measured $\phi=0.52$, we see discontinuous shear thickening, indicated by ${\rm d}\ln \eta / {\rm d} \ln \sigma > 1$ [dashed line]. Fitting $\eta_r^\infty(\phi)$ to Eq.~\ref{eq:KD}, Fig.~\ref{fig:PAAfc}~[inset (red)], gives the frictional jamming point $\phi_J= \phi_{\mu} = 0.49(1)$ [with $\ell = 2.2(2)$], the same as the high stress $\phi_J$ for bare calcite suspensions to within error. 


The absence of any observable shear thinning implies that $\sigma_a \to 0$. The observation of shear thickening means that, instead, $\sigma^*$ is now finite: a repulsive interaction must be overcome to press particles into mechanical contact, so that $\theta \to 0$. With this stabilising repulsive interaction, the viscosity should diverge only at random close packing at $\sigma \to 0$. Fitting $\eta_r^0(\phi)$ to Eq.~\ref{eq:KD}, Fig.~\ref{fig:PAAfc}~[inset (black)], we find that, indeed, $\phi_J = 0.62(3)$ [with $\ell = 2.2(4)$], consistent with the value of $\phi_J = 0.60$ determined separately using a powder compaction test (DS/EN 1097-4:2008). Using our two fitted values of $\phi_{\rm rcp}, \phi_\mu$ and $\ell = 2.2$, from the high-shear viscosity divergence, we fit Eq.~\ref{eq:f} to our data to find $\sigma^* = \SI{3.0(2)}{\pascal}$ (and $\beta = 1.02$). 

\begin{figure}
\includegraphics[width=\columnwidth]{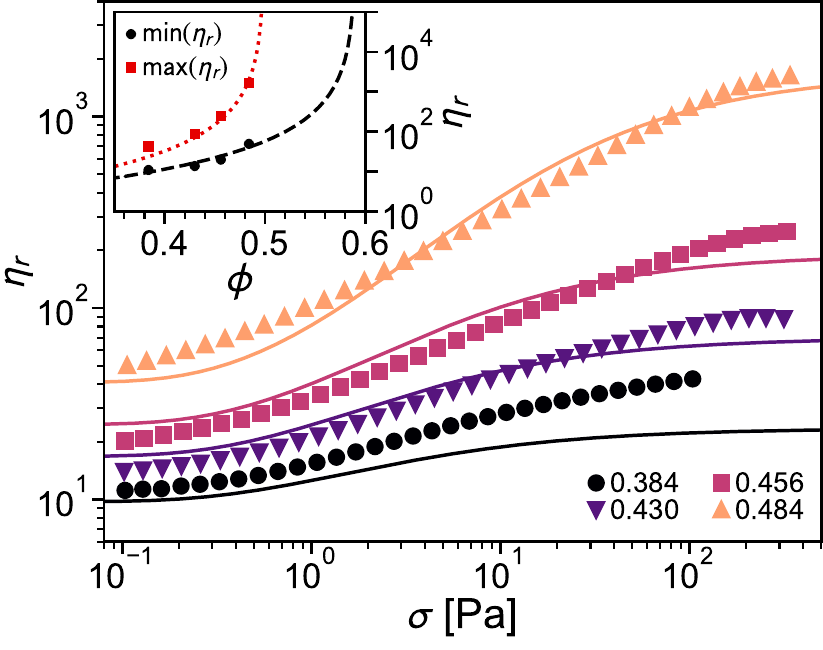}
\caption{Flow curves for alkyl-naphthalene sulphonate condensate (ANS) stabilised calcite suspensions. Symbols: relative viscosity vs stress, $\eta_r(\sigma)$, at volume fraction, $\phi$ (legend); lines, WC model ($\theta \to 0$) fit, $\sigma^* = \SI{1.0(1)}{\pascal}$ ($\beta = 0.67$). Inset: limiting $\phi_J$ used in WC model. Symbols: squares (red), $\max(\eta_r)$, high-shear viscosity; circles (black), $\min(\eta_r)$, low-shear viscosity. Lines: dotted (red), fit of Eq.~\ref{eq:KD} to $\max(\eta_r)$ with $\ell=2.2$ fixed from Fig.~\ref{fig:PAAfc} to find $\phi_{\rm J} = 0.499(3)$; dashed (black), fit to $\min(\eta_r)$ to find $\phi_{\rm J} = 0.59(1)$}
\label{fig:ANSfc}
\end{figure}

For ANS (in 85~wt.\% glycerol-water mixture, $\eta_s$ = \SI{110}{\milli \pascal \second}), a similar transition from a yield stress fluid to a shear-thickening suspension is seen, Fig.~\ref{fig:ANSfc}~(symbols). Repeating the same analysis procedure, we find $\phi_{\rm rcp} = 0.59(1)$ and $\phi_{\mu}=0.499(3)$, Fig.~\ref{fig:ANSfc}~(inset), consistent with the PAA data. Again, the flow curves suggest that $\sigma_a \rightarrow 0$, and fitting Eq.~\ref{eq:f} to the data now gives $\sigma^* = \SI{1.0(1)}{\pascal}$, so that, again $\theta \to 0$.

The effect of the third surfactant, PCE (in 50~wt.\% glycerol-water mixture, $\eta_s = \SI{6}{\milli \pascal \second}$), is similar at moderate to high stresses, $\sigma \gtrsim \SI{1}{\pascal}$, Fig.~\ref{fig:PCEfc}~(symbols). We see continuous shear thickening to a plateau at $\phi \leq 0.47$ and to fracture at higher concentrations. On the other hand, we now see shear thinning at $\sigma \lesssim \SI{1}{\pascal}$, with the appearance of a small yield stress at the highest solid volume fractions, $\sigma_y \approx \SI{0.2}{\pascal}$ at $\phi=0.51$. This value is, however, negligible compared to that of the bare calcite suspension at this volume fraction ($\gtrsim \SI{400}{\pascal}$). 

Fitting $\eta_r^\infty(\phi)$ to Eq.~\ref{eq:KD} gives same frictional jamming point, $\phi_{\mu} = 0.49(1)$ ($l=2.3$), as before. Fixing $\phi_{\rm rcp} = 0.62$, $\phi_{\rm alp} = 0.18$, $\beta=0.67$ and $\kappa=0.6$ from previous fittings, we find that the constraint model can give a reasonable account of the observed trends using $\sigma_a = \SI{0.3}{\pascal}$ and $\sigma^* = \SI{3}{\pascal}$, Fig.~\ref{fig:PCEfc} (dashed lines), so that $\theta \approx 0.1$. 


\section*{Discussion and conclusions\label{sec:discus}}

In the constraint rheology framework, the effect of adding surfactants is a matter of tuning the relative magnitudes of the two relevant characteristic stress scales, $\sigma^*$, beyond which sliding constraints form rapidly, and $\sigma_a$, beyond which rolling constraints break rapidly. The values of $(\sigma^*, \sigma_a)$ we have deduced, Table~\ref{tab:stresses}, should be taken as no more than order of magnitude estimates: varying the procedure used in inferring them from data would have given different results; but this does not alter the qualitative picture. 

\begin{figure}[t]
\includegraphics[width=\columnwidth]{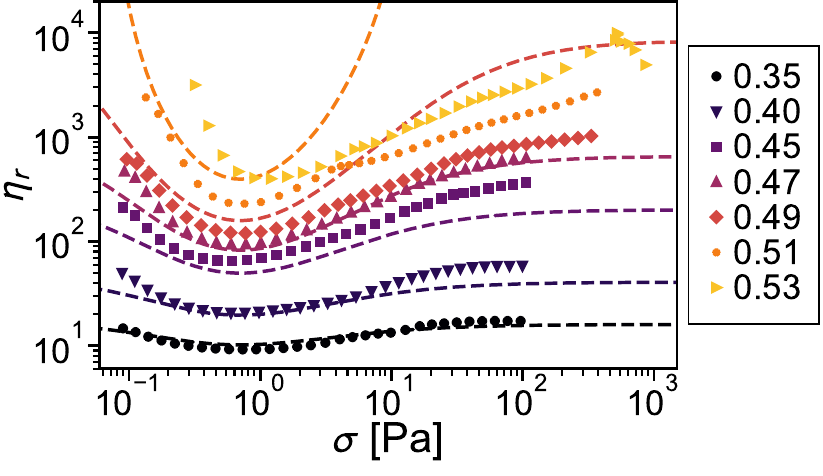}
\caption{Flow curves of polycarboxylate ether (PCE) stabilised ground calcite in a 50~wt.\% glycerol-water mixture. (a)~Relative viscosity vs stress, $\eta_r(\sigma)$. Symbols, data at volume fraction, $\phi$ (legend); dashed lines, Eqs.~\ref{eq:KD}-\ref{eq:phiJ2} with representative parameters, see text}
\label{fig:PCEfc}
\end{figure}

Dispersing bare nB calcite particles in a glycerol-water mixture gives suspensions in which a finite yield stress emerges at $\phi \gtrsim 0.18$ before jamming at all stresses at $\phi \gtrsim 0.5$, the frictional jamming point, $\phi_\mu$. The yield stress is due to adhesion of strength $\sigma_a \sim \SI{0.6}{\pascal}$ stabilising inter-particle frictional contacts, which form at infinitesimal applied stress, i.e., the onset stress $\sigma^*$ is vanishingly small. Adding any of the three surfactants confers a finite $\sigma^*$ of order $\sim 10^0 \, \si{\pascal}$. The difference between the surfactants is that while adding PAA and ANS reduces the adhesive strength to being immeasurably small, $\sigma_a \to 0$, adding PCE only decreases this stress scale by a factor of 2. Notably, this modest reduction in the adhesive strength still results in a large drop in the yield stress at, e.g., $\phi = 0.40$ from $\SI{4}{\pascal}$ to $\ll \SI{0.1}{\pascal}$, a factor of 40 (at least). With a large onset stress, $\theta \ll 1$, there is now no frictional contact network for adhesion to stabilise: frictional contacts are the primary determinants of the ability of the suspension to withstand finite applied stress. 

Our PCE data, Fig.~\ref{fig:PCEfc}, resemble those for hydrophobised glass spheres in water~\citep{brown2010generality}. Brown et al.~suggest that adding surfactant `eliminates \ldots clustering with its associated yield stress and reveals a region of underlying shear thickening.' Instead, our data indicate the converse: that surfactants reduce $\sigma_y$ in adhesive nB suspensions primarily by imparting a finite onset stress below which there can be no frictional network for inter-particle adhesion to stabilise.

Strikingly, the concentration at which the high-shear viscosity diverges for all three systems with surfactants occurs at $\phi \approx 0.50$ within experimental uncertainties, which is the frictional jamming point of the bare calcite suspension. At high applied stress, particles in the systems with surfactant additives interact as if they were bare. We interpret this as follows. Adding the amount of surfactant we used in each case provides approximately monolayer covering to the bare calcite particles providing a degree of steric stabilisation. Now, a finite stress, $\sigma^*$, is needed to push them into frictional contact. Thus, $\sigma^*$ is a measure of the stress needed to displace adsorbed surfactants from the calcite surface. Once the surfactant is displaced, the frictional interaction is again that between bare particles, accounting for the same $\phi_\mu$ in all four suspensions. 

In the case of PAA adsorbing on calcite, we can show quantitatively that this is a reasonable suggestion. The adsorption energy of \SI{2000}{\dalton} PAA (we used \SI{5100}{\dalton}) on calcite at room temperature is $E = \SI{15}{\kilo \joule \per \mole} \approx 6k_BT$ per polymer coil~\citep{sparks2015adsorption}. This allows us to estimate a local critical stress scale for desorption $\sigma^*_0 = E/R_g^3 \sim \SI{D6}{\pascal}$, which can be converted into the area of contact when friction is turned on, $r_0^2$, by equating the bulk and local forces, $\sigma^*_0 r_0^2 \sim \sigma^* d^2$, for particles of linear size $d$. This gives $r_0 \sim \SI{7}{\nano\metre}$, which is plausibly the length scale of surface roughness; so the picture is that $\sigma^*$ is the external stress needed to drive adsorbed PAA from asperities, exposing these to interact frictionally. This highlights the role of surfactant adsorption at the local level, as raised by \citet{mantellato2020shifting} for partial coverage of superplasticiser in cementitious suspensions.

\begin{table}
\centering
\begin{tabular}{|c|c|c|c|c|c|}
\hline
  Surfactant & \thead{$\sigma^*$ \\{[Pa]}} & \thead{$\sigma_a$\\{[Pa]}} & $\theta = \frac{\sigma_a}{\sigma^*}$ & \thead{$\sigma_y$ [Pa] at\\$\phi = 0.40$} & $\phi_\mu$\\
  \hline\hline
   None & $\to 0$ & 0.6 & $\to \infty $ & $4$ & 0.50(1)\\
   \hline
   PAA & 3 & $\to 0$ & $\to 0$ & $\to 0$ & 0.49(1)\\
   \hline
   ANS & 1 & $\to 0$ & $\to 0$ & $\to 0$ & 0.50\\
   \hline
   PCE & 3 & 0.3 & 0.1 & $\ll 0.1$ & 0.49 \\
   \hline
\end{tabular}
\caption{Estimates of characteristic stresses in calcite suspensions without and with surfactants. For each system: frictional onset stress, $\sigma^*$; adhesive strength, $\sigma_a$; stress scale ratio, $\theta$; yield stress, $\sigma_y$, at $\phi = 0.40$; and the fitted frictional jamming point, $\phi_\mu$}
\label{tab:stresses}
\end{table}

These findings lead to a new design principle for using surfactants to `tune' the rheology of nB adhesive suspensions. Given two surfactants that are equally effective as steric stabilisers, i.e., in lowering $\sigma_a$, the one that is more strongly adsorbed, i.e., with the higher adsorption energy $E$, should give a higher $\sigma^*$ and therefore be more effective in lowering $\sigma_y$. Indeed, in the limit of a high enough $\sigma^*$ relative to $\sigma_a$, adhesive bonds are all broken before any frictional contact network can be formed. The latter can therefore never be stabilised by adhesion, and a yield stress cannot emerge below $\phi_\mu$. Thus, $\sigma_y$ is reduced from some finite value to zero for all $\phi < \phi_\mu$, as is observed when we add PAA or ANS, Table~\ref{tab:stresses}. Such use of surfactants to confer a finite onset stress for nB adhesive suspensions to give essentially an infinite-fold reduction in the yield stress is perhaps the main, and certainly the most surprising, conclusion of this work, generalising a similar but less clear-cut finding in the use of surfactants to tune the rheology of molten chocolate~\citep{blanco2019conching}. 

\vspace{2em}
\noindent\textbf{Acknowledgements} The authors would like to thank Elena Blanco for providing the SEM micrograph.

\vspace{1em}
\noindent\textbf{Funding information} JAR~was funded by the UK Engineering and Physical Sciences Research Council (EPSRC) Centre for Doctoral Training in Soft Matter and Functional Interfaces (SOFI CDT) [Grant No.~EP/L015536/1]; and \mbox{AkzoNobel}, who also provided the surfactants used in this work. WCKP and REO'N were funded by the EPSRC [Grant No.~EP/N025318/1].

\vspace{1em}
\noindent\textbf{Data availability} The datasets generated during and/or analysed during the current study are available in the Edinburgh DataShare repository, \url{https://doi.org/10.7488/ds/2875}.
\vspace{1em}

\noindent\textbf{Conflict of interest} The authors declare that they have no competing interests.

\appendix

\section*{Appendix: Choosing surfactant concentrations}

\begin{figure}
\centering
\includegraphics[]{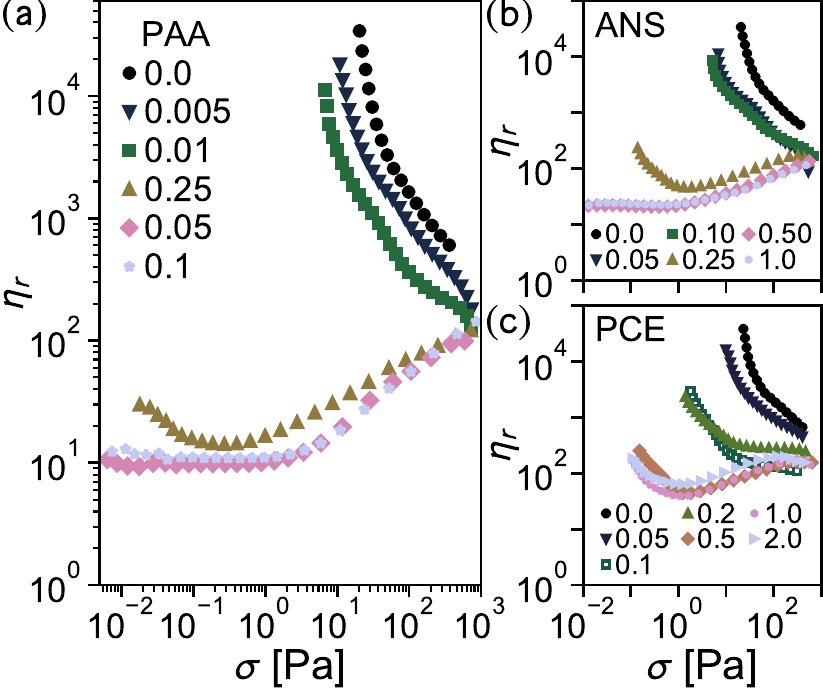}
\caption{Effect of surfactant concentration. (a)~Variation of polyacrylic acid (PAA) sodium salt. Symbols, relative viscosity as a function of stress, $\eta_r(\sigma)$, at a volume fraction, $\phi =0.44$, in a 50~wt.\%~glycerol-water mixture with PAA concentration weight percentage relative to calcite (w/w\%), legend. (b)~Alkyl-naphthalene sulphonate condensate sodium salt (ANS) variation. (c)~Polycarboxylate ether (PCE) variation}
\label{fig:sweep}
\end{figure}

Three surfactants, PAA, ANS and PCE, were used in this work to modify the interaction between calcite particles. We determined the concentration used for detailed investigation in each case by measuring the flow curve for a $\phi = 0.44$ suspension in a 50~wt\% glycerol-water mixture at increasing surfactant concentration until we reached the point where a further increase does not bring about further changes in the rheological behaviour. Measurements were made under imposed shear rate (TA Instruments ARES-G2, sandblasted plates) with a sweep at 6 points per decade from $\dot{\gamma}_{\min}=\SI{0.1}{\per \second}$ using a fixed step time of \SI{20}{\second} equilibration and \SI{10}{\second} of measurement. Note that as the PCE was supplied in a liquid form, we take the active component to be 50~wt.\% based on the product specification. The addition of surfactants at the concentrations used was found not to affect the glycerol-water mixture viscosities.

The data collected in this process are shown in Fig.~\ref{fig:sweep} for the three surfactants. For all cases we see a saturation in the effect of the surfactant, with no further change in the rheology measured after 0.05~w/w\% for PAA, 0.5~w/w\% for ANS and 1~w/w\% for PCE and the concentration at saturation is used for the experiments reported in the main text. [Note that a single anomalous flow curve was seen for 0.1~w/w\% PCE, Fig.~\ref{fig:sweep}(c); this is possibly due to sample under-filling.]

The amount of surfactant needed to achieve saturation effect can be understood quantitatively in the case of the PAA we used, which has a gyration radius of $R_g \approx \SI{3}{\nano\meter}$. It is easy to estimate that at 0.05~w/w\%, PAA coils of area $\sim R_g^2$ can cover $\approx \SI{0.5}{\meter^2\per\gram}$ of calcite surfaces. Separately, modelling our calcite particles as spheres with diameter $d = \SI{4}{\micro\meter}$, we estimate that the suspension has a specific surface area of $\approx \SI{0.5}{\meter^2\per\gram}$, although this is only an estimate due to asphericity and polydispersity. The saturation concentration therefore credibly represents complete monolayer coverage. We may surmise that the same may be true for the other two surfactants. This finding lends credence to our suggestion that the finite onset stress conferred by surfactants at the saturation concentration scales with the local stress needed to dislodge adsorbed surfactant molecules.


\begin{thebibliography}{38}
\providecommand{\natexlab}[1]{#1}
\providecommand{\url}[1]{{#1}}
\providecommand{\urlprefix}{URL }
\expandafter\ifx\csname urlstyle\endcsname\relax
  \providecommand{\doi}[1]{DOI~\discretionary{}{}{}#1}\else
  \providecommand{\doi}{DOI~\discretionary{}{}{}\begingroup
  \urlstyle{rm}\Url}\fi
\providecommand{\eprint}[2][]{\url{#2}}

\bibitem[{Al~Mahrouqi et~al.(2017)Al~Mahrouqi, Vinogradov, and
  Jackson}]{almahrouqi2017zeta}
Al~Mahrouqi D, Vinogradov J, Jackson MD (2017)
  \href{https://doi.org/10.1016/j.cis.2016.12.006}{Zeta potential of artificial
  and natural calcite in aqueous solution}. Adv Colloid Interface Sci 240:60 --
  76

\bibitem[{Barnes(1989)}]{barnes1989shear}
Barnes HA (1989) \href{https://doi.org/10.1122/1.4890747}{Shear-thickening
  (``dilatancy'') in suspensions of nonaggregating solid particles dispersed in
  {N}ewtonian liquids}. J Rheol 33(2):329--366

\bibitem[{Bergstr{\"o}m(1997)}]{bergstrom1997hamaker}
Bergstr{\"o}m L (1997)
  \href{https://doi.org/10.1016/S0001-8686(97)00003-1}{Hamaker constants of
  inorganic materials}. Adv Colloid Interface Sci 70:125 -- 169

\bibitem[{Blanco et~al.(2019)Blanco, Hodgson, Hermes, Besseling, Hunter,
  Chaikin, Cates, Van~Damme, and Poon}]{blanco2019conching}
Blanco E, Hodgson DJM, Hermes M, Besseling R, Hunter GL, Chaikin PM, Cates ME,
  Van~Damme I, Poon WCK (2019)
  \href{https://doi.org/10.1073/pnas.1901858116}{Conching chocolate is a
  prototypical transition from frictionally jammed solid to flowable suspension
  with maximal solid content}. Proc Natl Acad Sci (USA) 116:10303--10308

\bibitem[{Bonn et~al.(2017)Bonn, Denn, Berthier, Divoux, and
  Manneville}]{bonn2017yield}
Bonn D, Denn MM, Berthier L, Divoux T, Manneville S (2017)
  \href{https://doi.org/10.1103/RevModPhys.89.035005}{Yield stress materials in
  soft condensed matter}. Rev Mod Phys 89:035005

\bibitem[{Bossis et~al.(2017)Bossis, Boustingorry, Grasselli, Meunier, Morini,
  Zubarev, and Volkova}]{bossis2017discontinuous}
Bossis G, Boustingorry P, Grasselli Y, Meunier A, Morini R, Zubarev A, Volkova
  O (2017) \href{https://doi.org/10.1007/s00397-017-1005-4}{Discontinuous shear
  thickening in the presence of polymers adsorbed on the surface of calcium
  carbonate particles}. Rheol Acta 56(5):415--430

\bibitem[{Brown and Jaeger(2014)}]{brown2014shear}
Brown E, Jaeger HM (2014)
  \href{https://doi.org/10.1088/0034-4885/77/4/046602}{Shear thickening in
  concentrated suspensions: phenomenology, mechanisms and relations to
  jamming}. Rep Prog Phys 77(4):046602

\bibitem[{Brown et~al.(2010)Brown, Forman, Orellana, Zhang, Maynor, Betts,
  DeSimone, and Jaeger}]{brown2010generality}
Brown E, Forman NA, Orellana CS, Zhang H, Maynor BW, Betts DE, DeSimone JM,
  Jaeger HM (2010) \href{https://doi.org/10.1038/nmat2627}{Generality of shear
  thickening in dense suspensions}. Nat Mater 9(3):220

\bibitem[{Clavaud et~al.(2017)Clavaud, B{\'e}rut, Metzger, and
  Forterre}]{clavaud2017revealing}
Clavaud C, B{\'e}rut A, Metzger B, Forterre Y (2017)
  \href{https://doi.org/10.1073/pnas.1703926114}{Revealing the frictional
  transition in shear-thickening suspensions}. Proc Natl Acad Sci (USA)
  114(20):5147--5152

\bibitem[{Comtet et~al.(2017)Comtet, Chatt{\'e}, Nigu{\`e}s, Bocquet, Siria,
  and Colin}]{comtet2017pairwise}
Comtet J, Chatt{\'e} G, Nigu{\`e}s A, Bocquet L, Siria A, Colin A (2017)
  \href{https://doi.org/10.1038/ncomms15633}{Pairwise frictional profile
  between particles determines discontinuous shear thickening transition in
  non-colloidal suspensions}. Nat Commun 8:15633

\bibitem[{Dhar et~al.(2020)Dhar, Chattopadhyay, and
  Majumdar}]{dhar2020signature}
Dhar S, Chattopadhyay S, Majumdar S (2020)
  \href{https://doi.org/10.1088/1361-648X/ab5bd2}{Signature of jamming under
  steady shear in dense particulate suspensions}. J Phys: Condens Matter
  32(12):124002

\bibitem[{Eriksson et~al.(2007)Eriksson, Merta, and
  Rosenholm}]{eriksson2007calcite}
Eriksson R, Merta J, Rosenholm JB (2007)
  \href{https://doi.org/10.1016/j.jcis.2007.04.034}{The calcite/water
  interface: I. Surface charge in indifferent electrolyte media and the
  influence of low-molecular-weight polyelectrolyte}. J Colloid Interface Sci
  313(1):184 -- 193

\bibitem[{Estrada et~al.(2011)Estrada, Az\'ema, Radjai, and
  Taboada}]{estrada2011identification}
Estrada N, Az\'ema E, Radjai F, Taboada A (2011)
  \href{https://doi.org/10.1103/PhysRevE.84.011306}{Identification of rolling
  resistance as a shape parameter in sheared granular media}. Phys Rev E
  84:011306

\bibitem[{Guy et~al.(2015)Guy, Hermes, and Poon}]{guy2015towards}
Guy BM, Hermes M, Poon WCK (2015)
  \href{https://doi.org/10.1103/PhysRevLett.115.088304}{Towards a unified
  description of the rheology of hard-particle suspensions}. Phys Rev Lett
  115(8):088304

\bibitem[{Guy et~al.(2018)Guy, Richards, Hodgson, Blanco, and
  Poon}]{guy2018constraint}
Guy BM, Richards JA, Hodgson DJM, Blanco E, Poon WCK (2018)
  \href{https://doi.org/10.1103/PhysRevLett.121.128001}{Constraint-based
  approach to granular dispersion rheology}. Phys Rev Lett 121:128001

\bibitem[{Heim et~al.(1999)Heim, Blum, Preuss, and Butt}]{heim1999adhesion}
Heim LO, Blum J, Preuss M, Butt HJ (1999)
  \href{https://doi.org/10.1103/PhysRevLett.83.3328}{Adhesion and friction
  forces between spherical micrometer-sized particles}. Phys Rev Lett
  83:3328--3331

\bibitem[{Heymann et~al.(2002)Heymann, Peukert, and Aksel}]{heymann2002solid}
Heymann L, Peukert S, Aksel N (2002)
  \href{https://doi.org/10.1007/s00397-002-0227-1}{On the solid-liquid
  transition of concentrated suspensions in transient shear flow}. Rheol Acta
  41(4):307--315

\bibitem[{Hsu et~al.(2018)Hsu, Ramakrishna, Zanini, Spencer, and
  Isa}]{hsu2018roughness}
Hsu CP, Ramakrishna SN, Zanini M, Spencer ND, Isa L (2018)
  \href{https://doi.org/10.1073/pnas.1801066115}{Roughness-dependent tribology
  effects on discontinuous shear thickening}. Proc Natl Acad Sci USA
  115(20):5117--5122

\bibitem[{James et~al.(2018)James, Han, de~la Cruz, Jureller, and
  Jaeger}]{james2018interparticle}
James NM, Han E, de~la Cruz RAL, Jureller J, Jaeger HM (2018)
  \href{https://doi.org/10.1038/s41563-018-0175-5}{Interparticle hydrogen
  bonding can elicit shear jamming in dense suspensions}. Nat Mater
  17(11):965--970

\bibitem[{Krieger and Dougherty(1959)}]{krieger1959mechanism}
Krieger IM, Dougherty TJ (1959) \href{https://doi.org/10.1122/1.548848}{A
  mechanism for non‐Newtonian flow in suspensions of rigid spheres}. Transac
  Soc Rheol 3:137--152

\bibitem[{{KSL Staubtechnik GmbH}(2007)}]{eskal500data}
{KSL Staubtechnik GmbH} (2007) Eskal 500 datasheet

\bibitem[{Lerner et~al.(2012)Lerner, D{\"u}ring, and Wyart}]{lerner2012unified}
Lerner E, D{\"u}ring G, Wyart M (2012)
  \href{https://doi.org/10.1073/pnas.1120215109}{A unified framework for
  non-Brownian suspension flows and soft amorphous solids}. Proc Natl Acad Sci
  (USA) 109:4798--4803

\bibitem[{Lin et~al.(2015)Lin, Guy, Hermes, Ness, Sun, Poon, and
  Cohen}]{lin2015hydrodynamic}
Lin NYC, Guy BM, Hermes M, Ness C, Sun J, Poon WCK, Cohen I (2015)
  \href{https://doi.org/10.1103/PhysRevLett.115.228304}{Hydrodynamic and
  contact contributions to continuous shear thickening in colloidal
  suspensions}. Phys Rev Lett 115(22):228304

\bibitem[{Liu et~al.(2017)Liu, Jin, Chen, Makse, and Li}]{liu2017equation}
Liu W, Jin Y, Chen S, Makse HA, Li S (2017)
  \href{https://doi.org/10.1039/C6SM02216B}{Equation of state for random sphere
  packings with arbitrary adhesion and friction}. Soft Matter 13:421--427

\bibitem[{Mantellato and Flatt(2020)}]{mantellato2020shifting}
Mantellato S, Flatt RJ (2020)
  \href{https://doi.org/10.1111/jace.17040}{Shifting factor—A new paradigm
  for studying the rheology of cementitious suspensions}. J Am Ceram Soc
  103(6):3562--3574

\bibitem[{Mechtcherine et~al.(2020)Mechtcherine, Bos, Perrot, {da Silva},
  Nerella, Fataei, Wolfs, Sonebi, and Roussel}]{mechtcherine2020extrusion}
Mechtcherine V, Bos F, Perrot A, {da Silva} WL, Nerella V, Fataei S, Wolfs R,
  Sonebi M, Roussel N (2020)
  \href{https://doi.org/10.1016/j.cemconres.2020.106037}{Extrusion-based
  additive manufacturing with cement-based materials – Production steps,
  processes, and their underlying physics: A review}. Cem Concr Res 132:106037

\bibitem[{Mehta(1999)}]{mehta1999advancements}
Mehta PK (1999) Advancements in concrete technology. Concrete International
  21(6):69--76

\bibitem[{Piotte et~al.(1995)Piotte, Boss\'{a}nyi, Perreault, and
  Jolicoeur}]{piotte1995characterization}
Piotte M, Boss\'{a}nyi F, Perreault F, Jolicoeur C (1995)
  \href{https://doi.org/10.1016/0021-9673(95)00226-D}{Characterization of
  poly(naphthalenesulfonate) salts by ion-pair chromatography and
  ultrafiltration}. J Chromatogr A 704(2):377 -- 385

\bibitem[{Reith et~al.(2002)Reith, Müller, Müller-Plathe, and
  Wiegand}]{reith2002does}
Reith D, Müller B, Müller-Plathe F, Wiegand S (2002)
  \href{https://doi.org/10.1063/1.1471901}{How does the chain extension of poly
  (acrylic acid) scale in aqueous solution? A combined study with light
  scattering and computer simulation}. J Chem Phys 116(20):9100--9106

\bibitem[{Richards et~al.(2020)Richards, Guy, Blanco, Hermes, Poy, and
  Poon}]{richards2020role}
Richards JA, Guy BM, Blanco E, Hermes M, Poy G, Poon WCK (2020)
  \href{https://doi.org/10.1122/1.5132395}{The role of friction in the yielding
  of adhesive non-Brownian suspensions}. J Rheol 64(2):405--412

\bibitem[{Roussel(2007)}]{roussel2007rheology}
Roussel N (2007) \href{https://doi.org/10.1617/s11527-007-9313-2}{Rheology of
  fresh concrete: from measurements to predictions of casting processes}. Mater
  Struct 40(10):1001--1012

\bibitem[{Seto et~al.(2013)Seto, Mari, Morris, and
  Denn}]{seto2013discontinuous}
Seto R, Mari R, Morris JF, Denn MM (2013)
  \href{https://doi.org/10.1103/PhysRevLett.111.218301}{Discontinuous shear
  thickening of frictional hard-sphere suspensions}. Phys Rev Lett 111:218301

\bibitem[{Silbert(2010)}]{silbert2010jamming}
Silbert LE (2010) \href{https://doi.org/10.1039/C001973A}{Jamming of frictional
  spheres and random loose packing}. Soft Matter 6(13):2918--2924

\bibitem[{Sparks et~al.(2015)Sparks, Romero-González, El-Taboni, Freeman,
  Hall, Kakonyi, Swanson, Banwart, and Harding}]{sparks2015adsorption}
Sparks DJ, Romero-González ME, El-Taboni E, Freeman CL, Hall SA, Kakonyi G,
  Swanson L, Banwart SA, Harding JH (2015)
  \href{https://doi.org/10.1039/C5CP00945F}{Adsorption of poly acrylic acid
  onto the surface of calcite: an experimental and simulation study}. Phys Chem
  Chem Phys 17:27357--27365

\bibitem[{Wildemuth and Williams(1985)}]{wildemuth1985new}
Wildemuth CR, Williams MC (1985) \href{https://doi.org/10.1007/BF01329266}{A
  new interpretation of viscosity and yield stress in dense slurries: Coal and
  other irregular particles}. Rheol Acta 24(1):75--91

\bibitem[{Wilson and Davis(2000)}]{wilson2000viscosity}
Wilson HJ, Davis RH (2000) \href{https://doi.org/10.1017/S0022112000001695}{The
  viscosity of a dilute suspension of rough spheres}. J Fluid Mech 421:339--367

\bibitem[{Wyart and Cates(2014)}]{wyart2014discontinuous}
Wyart M, Cates ME (2014)
  \href{https://doi.org/10.1103/PhysRevLett.112.098302}{Discontinuous shear
  thickening without inertia in dense non-{Brownian} suspensions}. Phys Rev
  Lett 112(9):098302

\bibitem[{Zhou et~al.(1995)Zhou, Uhlherr, and Luo}]{zhou1995yield}
Zhou JZQ, Uhlherr PHT, Luo FT (1995)
  \href{https://doi.org/10.1007/BF00712315}{Yield stress and maximum packing
  fraction of concentrated suspensions}. Rheol Acta 34(6):544--561

\end{thebibliography}
\end{document}